\documentclass[twocolumn,prd,superscriptaddress, nofootinbib,showpacs,epsfig,preprintnumbers]{revtex4}
\usepackage{graphicx,epsfig,amsmath}

\newcommand{\comment}[1]{}

\newcommand{\lis}{\mbox{$^{7}$Li }}
\newcommand{\bes}{\mbox{$^7$Be }} 
\newcommand{\boe}{\mbox{$^8$B }}

\begin{document}

\flushbottom

\title{Constraining Big Bang lithium production with recent solar neutrino data}

\author{Marcell P. Tak\'acs}
\affiliation{Helmholtz-Zentrum Dresden--Rossendorf (HZDR), 01328 Dresden, Germany}
\affiliation{Technische Universit\"at Dresden,  01069 Dresden, Germany}

\author{Daniel Bemmerer}
\affiliation{Helmholtz-Zentrum Dresden--Rossendorf (HZDR), 01328 Dresden, Germany}

\author{Tam\'as Sz\"{u}cs}
\affiliation{Helmholtz-Zentrum Dresden--Rossendorf (HZDR), 01328 Dresden, Germany}

\author{Kai Zuber}
\affiliation{Technische Universit\"at Dresden,  01069 Dresden, Germany}

\begin{abstract}
The $^3$He($\alpha,\gamma$)$^7$Be reaction affects not only the production of $^7$Li in Big Bang nucleosynthesis, but also the fluxes of $^7$Be and $^8$B neutrinos from the Sun. This double role is exploited here to constrain the former by the latter. 
A number of recent experiments on $^3$He($\alpha,\gamma$)$^7$Be provide precise cross section data at $E$ = 0.5-1.0 MeV center-of-mass energy. However, there is a scarcity of precise data at Big Bang energies, 0.1-0.5\,MeV, and below. This problem can be alleviated, based on precisely calibrated $^7$Be and $^8$B neutrino fluxes from the Sun that are now available, assuming the neutrino flavour oscillation framework to be correct. These fluxes and the standard solar model are used here to determine the $^3$He($\alpha,\gamma$)$^7$Be astrophysical S-factor at the solar Gamow peak, $S_{34}^{\nu}(23^{+6}_{-5}\,\text{keV})$ = 0.548$\pm$0.054~keV\,b.
This new data point is then included in a re-evaluation of the $^3$He($\alpha,\gamma$)$^7$Be S-factor at Big Bang energies, following an approach recently developed for this reaction in the context of solar fusion studies. The re-evaluated S-factor curve is then used to re-determine the $^3$He($\alpha,\gamma$)$^7$Be thermonuclear reaction rate at Big Bang energies. The predicted primordial lithium abundance is $^7{\rm Li/H}$ = 5.0 $\times 10^{-10}$, far higher than the Spite plateau. 
\end{abstract}
\preprint{arXiv version}

\maketitle

\section{Introduction}
\label{sec:Introduction}

The prediction of the light element abundances in Big Bang nucleosynthesis (BBN) is a pillar of modern cosmology. The consistent description of abundances over ten orders of magnitude can be considered as a big success. Latest data on the cosmic microwave background obtained by the Planck mission
fix the baryon density and thus the baryon-photon ratio $\eta$ \cite{Planck14-Cosmology}. However, there is still a puzzling disagreement between
the observed abundance of $^7$Li in metal poor stars of $^7$Li/H = (1.6$\pm$0.3)$\times$10$^{-10}$ \cite{PDG14} and the prediction from BBN of $^7$Li/H = (4.95$\pm$0.39)$\times$10$^{-10}$ \cite{Coc14-JCAP}. For a recent review of the lithium problem, see Ref.~\cite{Fields11-ARNPS}. 

The production of \lis in BBN depends on thermonuclear reaction rates $N_{\rm A}\langle\sigma v\rangle$, in particular that of the $^3$He($\alpha,\gamma$)$^7$Be reaction called hereafter $R_{34}$. The thermonuclear reaction rate $R_{34}$, in turn, depends on the $^3$He($\alpha,\gamma$)$^7$Be cross section $\sigma_{34}(E)$ and on the temperature $T$ prevalent in the astrophysical scenario under study:
\begin{multline} 
R_{34} \equiv N_{\rm A}\langle\sigma v\rangle_{34}(T) = \\ 
N_{\rm A} \frac{(8/\pi)^{1/2}}{\mu^{1/2}(k_{\rm B}T)^{3/2}} \int \limits_0^\infty E \sigma_{34}(E) \exp\left(-\frac{E}{k_{\rm B}T}\right) dE \label{eq:tnrr}
\end{multline}
with $E$ the center-of-mass energy and $\mu=m_3m_4/(m_3+m_4)$ the reduced mass of the two reaction partners $^3$He and $^4$He.

At astrophysical energies, the cross section $\sigma_{34}(E)$ exhibits an exponential-like energy dependence and can be parameterized as the astrophysical S-factor $S_{34}(E)$ which varies only very slowly with energy in the case of $^3$He($\alpha,\gamma$)$^7$Be \cite{Adelberger11-RMP}. The S-factor is defined by the following equation:
\begin{equation} \label{eq:sfactor}
\sigma_{34}(E) = \frac{1}{E} S_{34}(E) \exp\left[-2 \pi Z_1 Z_2 \frac{e^2}{\hbar} \sqrt{\frac{\mu}{2 E}} \right]
\end{equation}
where $Z_1 Z_2 e^2$ is the product of the nuclear charges of the two reacting nuclei. Inserting Eq.~(\ref{eq:sfactor}) in Eq.~(\ref{eq:tnrr}), it follows:
\begin{multline} \label{eq:tnrr2}
R_{34} \propto
\int \limits_0^\infty S_{34}(E) \exp\left[-\frac{E}{k_{\rm B}T} - 2 \pi Z_1 Z_2 \frac{e^2}{\hbar} \sqrt{\frac{\mu}{2 E}} \right] dE
\end{multline}
The exponential term is the so-called Gamow peak. The first term inside the exponential function forms the high-energy edge of the Gamow peak, given by the exponential decrease of the Maxwell-Boltzmann energy distribution. The second term forms the low-energy edge, given by the exponential-like decrease of the cross section. The energy range of this peak indicates where $S_{34}(E)$ must be integrated in order to obtain the thermonuclear reaction rate. 

\begin{figure}
\centering
\includegraphics[width=\linewidth]{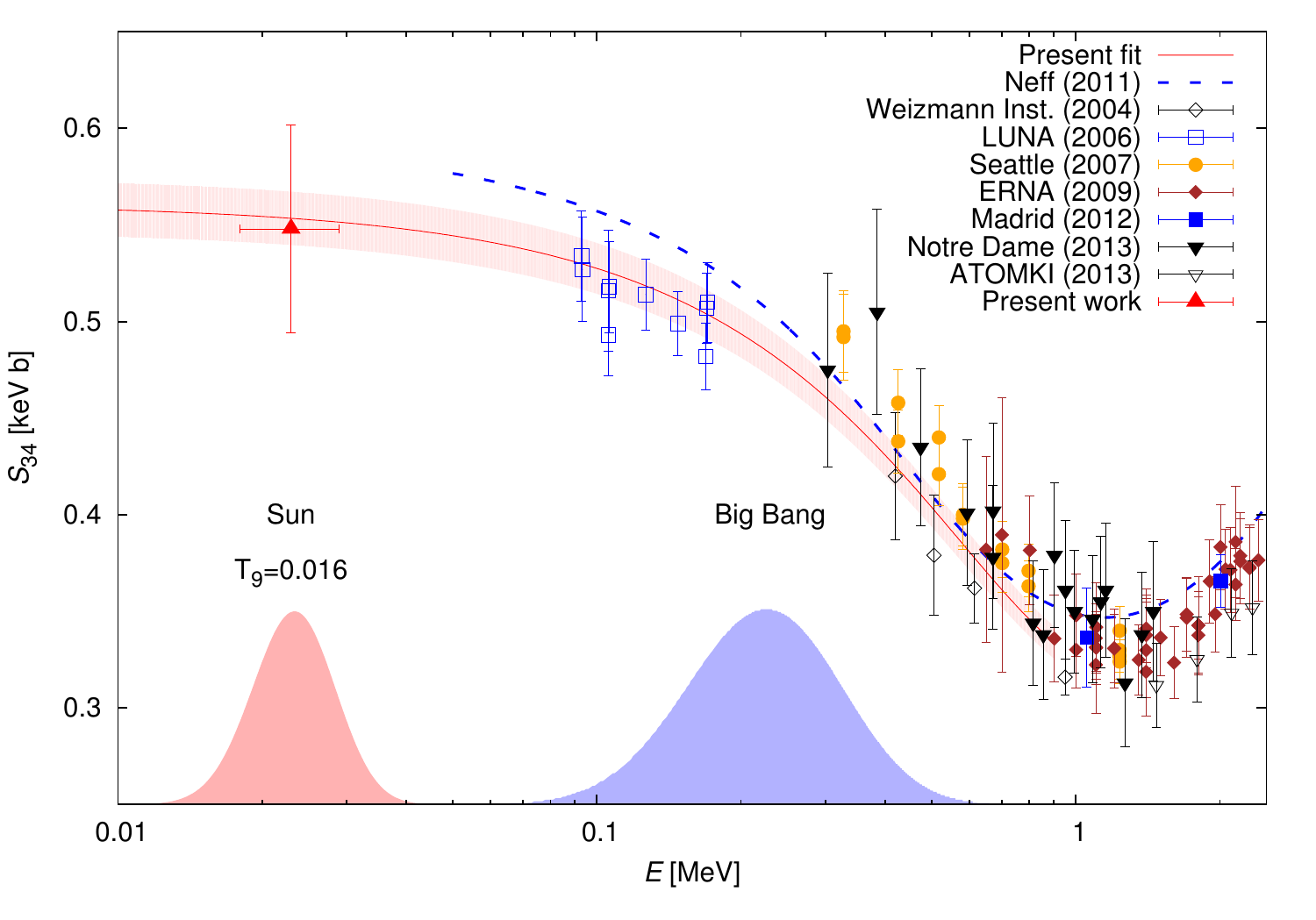}
\caption{Cross section of the $^3$He($\alpha$,$\gamma$)$^7$Be reaction, parameterized as the astrophysical $S$-factor. The present new data point (sec.~\ref{sec:sfactor}) is plotted together with previous experimental data \cite{NaraSingh04-PRL,Bemmerer06-PRL,Brown07-PRC,DiLeva09-PRL,Carmona12-PRC,Bordeanu13-NPA,Kontos13-PRC}. The previous theoretical curve (dashed blue curve, \cite{Neff11-PRL}), and the present new extrapolation (red curve, shaded area for the uncertainty) are shown. The solar Gamow peak and the relevant energy range for BBN (see text) are displayed at the lower end of the plot.}
\label{fig:sfactor}
\end{figure}

In the case of BBN, the S-factor must be known over a wide range in center-of-mass energies $E$. This range is estimated here by measuring the effect of a small change in the assumed S-factor at one given energy on the final $^7$Li abundance at the end of BBN, following the approach of Nollett and Burles \cite{Nollett00-PRD}. The relevant energy range is found to be $E$ = 0.1-0.5 MeV (Fig.~\ref{fig:sfactor}), consistent with the previous result by Ref.~\cite{Nollett00-PRD}. Subsequently, also the relevant temperature range for $^7$Be production in BBN is determined by arbitrarily setting $R_{34}$ to zero above a certain temperature, resulting in $T_9$=0.30-0.65, if a relevant effect is defined as a 2.5\% contribution on the $^7$Be yield. When converting these temperatures to Gamow energies, the resultant relevant energy range is consistent with the one based on the Nollett and Burles \cite{Nollett00-PRD} approach, adopted here.

A number of recent $S_{34}(E)$ determinations are available at $E$ $>$ 0.3\,MeV \cite{NaraSingh04-PRL,Brown07-PRC,DiLeva09-PRL,Carmona12-PRC,Bordeanu13-NPA,Kontos13-PRC}, allowing to form a weighted average and judge the precision of the recommended value (Fig.~\ref{fig:sfactor}). However, this abundance of recent experimental data covers only the upper third of the relevant energy range. At lower energy, the exceedingly low cross section is a challenge for experimentalists. As a consequence, recent data for $E$ $\leq$ 0.3\,MeV are available only from one experiment \cite{Bemmerer06-PRL}, performed at the LUNA accelerator deep underground in the Gran Sasso laboratory, Italy. 

It should be noted that $S_{34}(E)$ data reported in the period from the 1950s to the 1980s \cite{Holmgren59-PR,Parker63-PR,Nagatani69-NPA,Kraewinkel82-ZPA,Osborne82-PRL,Alexander84-NPA,Hilgemeier88-ZPA} are omitted from the present discussion, following the approach of a recent review \cite{Adelberger11-RMP}. These data \cite{Holmgren59-PR,Parker63-PR,Nagatani69-NPA,Kraewinkel82-ZPA,Osborne82-PRL,Alexander84-NPA,Hilgemeier88-ZPA} are usually less well documented than the more recent works \cite{NaraSingh04-PRL,Bemmerer06-PRL,Brown07-PRC,DiLeva09-PRL,Carmona12-PRC,Bordeanu13-NPA,Kontos13-PRC} and have larger error bars. 

The scarcity of recent low-energy $S_{34}(E)$ is addressed here based on the fact that actually the Gamow peak is rather narrow for low temperatures (see the solar Gamow peak in Fig.~\ref{fig:sfactor}). 
Here, $S_{34}(E_{\rm Gamow}^{\rm Sun})$ is determined from $N_{\rm A}\langle\sigma v\rangle_{34}(T^{\rm Sun})$. The latest solar neutrino and cosmological data are used. The additional low-energy data point is used to re-determine the primordial lithium abundance.

A related idea has previously been explored by Cyburt et al. a decade ago \cite{Cyburt04-PRD-7Li}. That work was based on the neutrino data available at the time from the Sudbury Neutrino Observatory (SNO), and on the WMAP cosmological survey. 

The present work uses newly available cross section, solar neutrino, microwave background, and neutron lifetime data, which are summarized in sec.\,\ref{sec:InputData}. Using an approach and errors described in sections \ref{sec:Approach} and \ref{sec:ErrorAnalysis}, respectively, $S_{34}(E_{\rm Gamow}^{\rm Sun})$ is determined, limiting the use of the solar neutrino data to its strict range of applicability, the temperature range of the solar Gamow peak  (sec.\,\ref{subsec:sfactor_solar}). Subsequently, the new data point is included in a re-evaluation of the $^3$He($\alpha$,$\gamma$)$^7$Be S-factor at Big Bang energies (sec.\,\ref{subsec:sfactor_bbn}). The predicted lithium abundance from BBN is subsequently updated (sec.\,\ref{sec:ReactionRate}), and a summary and outlook are given (sec.\,\ref{sec:Summary}). In the appendix, the reaction rate is given both in parameterized and in tabular forms.

\section{Input data}
\label{sec:InputData}

The Sudbury Neutrino Observatory (SNO) reports a $^8$B solar neutrino flux of
\begin{equation*}
\phi^{\text{exp}}_{\text{B}} = 5.25 \pm 0.16 \text{ (stat)} ^{+0.11}_{-0.13} \text{ (sys)} \times 10^6 {\rm cm}^{-2}{\rm s}^{-1}
\end{equation*}
taking into account the loss in the amount of electron neutrinos due to the mixing among the neutrino families \cite{SNO13-PRC_all3phases}. This is equivalent to 3.9\% precision (systematical and statistical uncertainties combined in quadrature) and consistent with the determination made by Super-Kamiokande \cite{SuperK13-NPBPS}. 

The flux of $\bes$  neutrinos was measured by BOREXINO \cite{Borexino14-PRD}, resulting in a value of  
\begin{equation*}
\phi^{\text{exp}}_{\text{Be}} = 4.75 ^{+0.26}_{-0.22} \times 10^9 {\rm cm}^{-2}{\rm s}^{-1}
\end{equation*}
with 5.5\% total uncertainty.

The value for the baryonic density found by the Planck mission \cite{Planck14-Cosmology} is 
\begin{equation*}
\Omega_bh^2 = 0.02205 \pm 0.00028
\end{equation*}

This parameter is an important input for BBN calculations, in addition to the thermonuclear reaction rates of the relevant nuclear reactions. The lifetime of the neutron has only a weak effect on Big Bang $^7$Li. For the present work, the recently recommended value of $\tau_{\rm n}=880.3 \pm 1.1 \text{ s}$ \cite{PDG14} is used for consistency. However, different values from 878-885~s change the final $^7$Li abundance only slightly.

\section{Description of the approach}
\label{sec:Approach}

In this work, no solar model calculations are performed. Instead, the so-called standard solar model developed by John Bahcall and co-workers is used, hereafter called SSM. The partial derivatives for the various SSM input parameters are available in tabulated form in the most recent SSM publication by Serenelli {\it et al.} \cite{Serenelli13-PRD}. Henceforth, the terminology and numbers from this work are used.

The SSM uses a number of input parameters, including the solar age, luminosity, opacity, diffusion rate, the key thermonuclear reaction rates (herein called $R_i$, where $i$ denotes the nuclear reaction under study), and the zero-age abundance of important elements (He, C, N, O, Ne, Mg, Si, S, Ar, Fe). A change in one or several of these input parameters may cause a change in the predicted neutrino fluxes. The sensitivity of flux $\phi_{i}$ for a variation in an arbitrary parameter $\beta_{j}$ can be expressed by the logarithmic partial derivatives $\alpha(i,j)$ given by the following relation:
\begin{equation} \label{eq:Def_alpha}
\alpha(i,j)=\frac{\partial ln[\phi_{i}/\phi^{\rm SSM}_{i}]}{\partial ln[\beta_{j}/\beta^{\rm SSM}_{j}]}
\end{equation}
where $\phi^{\rm SSM}_{i}$ and $\beta^{\rm SSM}_{j}$ represent the best theoretical values from the SSM. In the present work, the derivatives from Ref.~\cite{Serenelli13-PRD} are used (Table~\ref{tab:derivatives}). The above defined logarithmic partial derivatives can be used to approximate relatively small changes in the neutrino flux as a simple power law:
\begin{equation} \label{eq:PowerLaw}
\frac{\phi_{i}}{\phi^{\rm SSM}_{i}}=\prod \limits_{j} \limits^{N} \Big(\frac{\beta_{j}}{\beta^{\rm SSM}_{j}}\Big) ^{\alpha(i,j)}
\end{equation}

The parameter of interest in the present work is the $S$-factor of  the $^3$He($\alpha,\gamma$)$^7$Be reaction, here denoted as $S_{34}$. This nuclear reaction is located at the beginning of the pp-2 and pp-3 branches of the pp-chain of hydrogen burning, and thus the value of $S_{34}$ strongly affects the $^7$Be and $^8$B neutrino fluxes, which is reflected in partial derivatives that are close to unity: $\alpha(Be,S_{34}) \approx \alpha(B,S_{34}) \approx 0.8$.

Now, by fixing all parameters except for $R_{34}$ at their SSM best-fit value, Eq.~(\ref{eq:PowerLaw}) is shortened to:
\begin{equation}\label{eq:ShortPowerLaw}
\frac{\phi^{\text{exp}}_{\text{Be}}}{\phi^{\text{SSM}}_{\text{Be}}} = \left(\frac{R^{\nu,\rm Be}_{34}}{R^{\rm SSM}_{34}}\right)^{\alpha(\text{Be},S_{34}^{\text{SSM}})}
\end{equation}
when using the experimental flux of $^7$Be neutrinos $\phi^{\text{exp}}_{\text{Be}}$. An analogous relation is obtained based on the $^8$B neutrino flux $\phi^{\text{exp}}_{\text{B}}$. Both numbers can be found in sec.~\ref{sec:InputData}. 

Solving for the thermonuclear reaction rate $R^{\nu,\rm Be/B}_{34}$, the following relations are obtained:
\begin{eqnarray}
R_{34}^{\nu,\text{Be}} & = & \Big(\frac{\phi^{\text{exp}}_{\text{Be}}}{\phi^{\text{SSM}}_{\text{Be}}}\Big)^{\alpha^{-1}(\text{Be},S_{34})}R_{34}^{\text{SSM}} \label{eq:R34A} \\
R_{34}^{\nu,\text{B}} & = & \Big(\frac{\phi^{\text{exp}}_{\text{B}}}{\phi^{\text{SSM}}_{\text{B}}}\Big)^{\alpha^{-1}(\text{B},S_{34})}R_{34}^{\text{SSM}} \label{eq:R34B}
\end{eqnarray}
The nuclear reaction rate $R_{34}$ used for Equations~(\ref{eq:ShortPowerLaw}-\ref{eq:R34B}) applies to a certain range of temperatures. The emission of $^7$Be neutrinos is known to originate from a narrow burning zone at the center of the Sun, at radii below 0.15$R_{\odot}$ (where $R_{\odot}$ is the the solar radius), with a temperature $T_9$ = 0.011-0.016, close to the nominal central temperature. The $^8$B neutrino emission originates from an even narrower burning zone, below 0.10$R_{\odot}$. Therefore, it can be assumed that to good approximation the relevant temperature for the $^3$He($\alpha$,$\gamma$)$^7$Be reaction is the central temperature of the Sun, $T_9$ = 0.016. 
Thus, equations~(\ref{eq:ShortPowerLaw}-\ref{eq:R34B}) apply to the nuclear reaction rate in the energy range of the solar Gamow peak (fig.~\ref{fig:sfactor}). The value of the reaction rate at energies that lie outside the Gamow peak does not affect solar fusion.

\begin{table}[t]
\begin{center}
{\small
\begin{tabular}{lrcrc}
\hline\noalign{\smallskip}
Parameter & $\alpha(\text{Be},j)$&$\frac{\Delta\phi_{\text{Be}}}{\phi_{\text{Be}}}(\%)$&$\alpha(\text{B},j)$&$\frac{\Delta\phi_{\text{B}}}{\phi_{\text{B}}}(\%)$ \\
\noalign{\smallskip}
\hline
Luminosity & 3.434 & 1.4 & 6.914 & 2.8\\
Opacity & 1.210 & 3.0 & 2.611 & 6.5\\
Age & 0.760 & 0.3 & 1.345 & 0.6\\
Diffusion & 0.126 & 1.9 & 0.267 & 4.0\\
$R_{11}$ - p+p & -1.024 & 1.0 & -2.651 & 2.6\\
$R_{33}$ - $^3$He +$^3$He & -0.428 & 2.2 & -0.405 & 2.1\\
$R_{34}$ - $^3$He +$^4$He & 0.853 & (4.6) & 0.806 & (4.3)\\
$R_{17}$ - p +$^7$Be & - & - & 1.000 & 7.7\\
$R_{e7}$ - e +$^7$Be & - & - & -1.000 & 2.0\\
Composition* & -& 4.6 & - & 9.7\\
\hline
Total uncertainty & & 6.5& & 15.3 \\
\hline
\end{tabular}
}
\end{center}
\caption{\label{tab:derivatives}
Logarithmic partial derivatives $\alpha(\text{Be},j)$ and $\alpha(\text{B},j)$, as defined by Eq.~(\ref{eq:Def_alpha}) and their contributions to the total uncertainty of the predicted SSM flux. Values and uncertainties are taken from \cite{Serenelli13-PRD}, except for the solar composition (see text). See sec.~\ref{sec:ErrorAnalysis} for details.}
\end{table}

\section{Error analysis}
\label{sec:ErrorAnalysis}

Table~\ref{tab:derivatives} lists the most important logarithmic partial derivatives $\alpha(\text{Be/B},j)$ discussed here. In addition, the Table lists the contribution of each parameter to the SSM error budget. Values and errors are taken from the most recent SSM paper by Serenelli {\it et al.} \cite{Serenelli13-PRD}. Two parameters merit a more detailed discussion:

First, the elemental composition of the Sun. It has undergone a significant revision from the GS98 \cite{Grevesse98-SSR} to the AGSS09 \cite{Asplund09-ARAA} abundance compilations. The determination of the abundance of a given element requires the modelling of the related absorption lines in the solar spectrum thus modelling the solar atmosphere. In the time interval from 1998 to 2005/2009, the modeling of the solar atmosphere was updated from a one-dimensional, time-independent, hydrostatic \cite{Grevesse98-SSR} to a three-dimensional, time-dependent hydrodynamical model \cite{Asplund09-ARAA}. 

The adoption of three-dimensional modeling in AGSS09 led to a significant downward reduction of the abundances of the so-called "metals", the name given in solar physics to all elements that are heavier than helium. The mass fraction $Z$ for "metals" in the Sun changed from 0.0169 \cite{Grevesse98-SSR} to 0.0134 \cite{Asplund09-ARAA}. The carbon and nitrogen abundances decreased by 19\%, and the oxygen abundance even by 28\% from GS98 to AGSS09. 

These significant revisions in the abundances of important elements lead to a contradiction between SSM predictions and helioseismological observations \cite{Serenelli09-ApJL}, when the new abundances are incorporated in the SSM. For the present purposes, the problem of the elemental abundances must be set aside. This is accomplished by adopting the average of the two different SSM predictions (the first one based on GS98, the second one based on AGSS09) as value and half the difference as uncertainty (Table~\ref{tab:fluxes}). In this manner, within their error bars the present conclusions apply to both the GS98 and AGSS09 elemental abundances. 

\begin{table}[tb]
\begin{center}
{\small
\begin{tabular}{llll}
\hline\noalign{\smallskip}
Elemental comp.  & $\phi(\bes)$&$\phi(\boe)$&Ref. \\
\noalign{\smallskip}
\hline
GS98 \cite{Grevesse98-SSR} & 5.00 & 5.58 &\cite{Serenelli11-ApJ}\\
AGSS09 \cite{Asplund09-ARAA} & 4.56 & 4.59 & \cite{Serenelli11-ApJ}\\
Average & 4.78 $\pm$ 0.22 & 5.09 $\pm$ 0.49 & This work\\
\hline
\end{tabular}
}
\end{center}
\caption{\label{tab:fluxes}
Predicted solar neutrino fluxes from the SSM for two different elemental abundances, taken from \cite{Serenelli13-PRD}. The average adopted here includes both results with its error bar.}
\end{table} 

Second, the astrophysical reaction rate of the $^3$He($\alpha$,$\gamma$)$^7$Be reaction, $R_{34}$. The value of $R_{34}$ taken in the SSM calculations followed here \cite{Serenelli13-PRD} is the recommended curve by the "Solar Fusion cross sections II" review \cite{Adelberger11-RMP}. However, in order to avoid double counting, the uncertainty of $R_{34}$ is left out when computing the total uncertainty (Table~\ref{tab:derivatives}). Instead, this parameter and its uncertainty are re-determined here based on all the other parameters. 

With these two modifications, the total uncertainty of the flux prediction is 6.5\% for $\phi_{\rm Be}^{\rm SSM}$ and 15.3\% for $\phi_{\rm B}^{\rm SSM}$. If one were to select just one of the two solar elemental compositions and its uncertainty, the total error budget would decrease to  4.5\% and 11.9\%, respectively.

The thermonuclear reaction rate $R_{34}$ is directly proportional to the astrophysical S-factor $S_{34}$ (Eq.~\ref{eq:tnrr2}) in the relevant energy range. Therefore, the relative errors derived for $R_{34}$ have to be used also for $S_{34}$.

\section{S-factor result}
\label{sec:sfactor}

\subsection{Determination of $S_{34}$ at the solar Gamow peak}
\label{subsec:sfactor_solar}

Using Eqns.~(\ref{eq:R34A}, \ref{eq:R34B}), the astrophysical S-factor is now determined here. For $R_{34}^{\rm SSM}$, the "Solar Fusion II" S-factor parameterization \cite{Adelberger11-RMP} has been used, therefore the new S-factor is found by rescaling the value of this parameterization at the solar Gamow peak energy:
\begin{eqnarray}
S_{34}^{^7\text{Be}}(23^{+6}_{-5}\,\text{keV}) &=& 0.548 \pm 0.054 \text{ keV b} \\ 
S_{34}^{^8\text{B}}(23^{+6}_{-5}\,\text{keV}) &=& 0.58 \pm 0.11 \text{ keV b}
\end{eqnarray}
The two data points are in good agreement with each other. Most of the contributions to the error budget that are common to both data points are from factors such as the elemental abundances that affect both the $^7$Be and $^8$B fluxes in the same direction, and  at the same time affect the $^8$B-based result more strongly than the $^7$Be-based one. Therefore, an averaging of the two numbers actually leads to a higher total uncertainty than the error bar of the $^7$Be-based value. Therefore, $S_{34}^{^7\text{Be}}(23^{+6}_{-5}\,\text{keV})$ is adopted as the final result here. 

The $S_{34}^{^7\text{Be}}(23^{+6}_{-5}\,\text{keV})$ value confirms that the shape of the "Solar Fusion II" recommended S-factor curve is correct at low energy (Fig.~\ref{fig:sfactor}). The present new value cannot be directly compared to the theory curve by Neff, which does not extend to such low energies for numerical reasons \cite{Neff11-PRL}. 

\begin{table}[t]
\begin{center}
{\small
\begin{tabular}{lr@{$\pm$}lc}
\hline\noalign{\smallskip}
Reference & \multicolumn{2}{c}{$S_{34}(0)$ [keV b]} & Inflation factor \\
\noalign{\smallskip}
\hline
Weizmann \cite{NaraSingh04-PRL} & 0.538 & 0.015 & 1.00\\
LUNA \cite{Bemmerer06-PRL,Gyurky07-PRC,Confortola07-PRC} & 0.550 & 0.017 & 1.06\\
Seattle \cite{Brown07-PRC}& 0.598 & 0.019 & 1.15\\
ERNA \cite{DiLeva09-PRL} & 0.582 & 0.029 & 1.03\\
Notre Dame \cite{Kontos13-PRC} & 0.593 & 0.048 & 1.00\\
Present work & 0.556 & 0.055 & 1.00\\
\hline \hline
Combined result & 0.561 & 0.011 & 1.32 \\
\hline
\end{tabular}
}
\end{center}
\caption{\label{tab:S340}
Determination of $S_{34}(0)$ from recent experimental data, using Eq.~(\ref{eq:Adelfit}) as fit function. See text for details.}
\end{table} 

\subsection{Combined fit of $S_{34}$ for BBN purposes}
\label{subsec:sfactor_bbn}

As a next step, the combined analysis of all experimental data points is carried out, repeating the approach of "Solar Fusion II" but adding the present new neutrino-based data point and the new data set from Notre Dame that became available in the meantime \cite{Kontos13-PRC}. The same analytical function as in "Solar Fusion II" is again used here, namely
\begin{multline} 
\label{eq:Adelfit}
S_{34}(E) = S_{34}(0) \exp(-0.580E) \\
(1-0.4054E^2+0.577E^3-0.1353E^4)
\end{multline}
The curve is based on the microscopic model by Nollett (Kim A potential) \cite{Nollett01-PRC} and was already previously used for fitting the experimental data \cite{Adelberger11-RMP}. A previous similar fit with the alternative microscopic model by Kajino \cite{Kajino86-NPA} gave consistent results. See Ref.~\cite{Adelberger11-RMP} for more details on those two models and the fitting approach. In the present work, only eq.~(\ref{eq:Adelfit}), based on Ref.~\cite{Nollett01-PRC}, is used. All the experimental data \cite[present]{NaraSingh04-PRL,Bemmerer06-PRL,Brown07-PRC,DiLeva09-PRL,Kontos13-PRC} lie near this curve (fig.~\ref{fig:sfactor}). 

For the analysis, each experimental data set \cite[present]{NaraSingh04-PRL,Bemmerer06-PRL,Brown07-PRC,DiLeva09-PRL,Kontos13-PRC} is fitted with the analytical function (\ref{eq:Adelfit})
in the energy range 0$\leq E \leq$1.002~MeV, and a value of $S_{34}(0)$ is then found for this particular data set. The data from Madrid and from ATOMKI \cite{Carmona12-PRC,Bordeanu13-NPA} are excluded, because for those two cases all of the data points fall outside the energy range of applicability of Eq.~(\ref{eq:Adelfit}). However, these data \cite{Carmona12-PRC,Bordeanu13-NPA} are in good agreement with other data sets which include data points both in the Madrid/ATOMKI energy range and in the range of applicability of the fit \cite{DiLeva09-PRL,Kontos13-PRC}. Therefore, no bias is introduced by the necessary omission of Refs.~\cite{Carmona12-PRC,Bordeanu13-NPA}. For each fitted data set, an inflation factor is determined from the goodness of the fit to the data, again following Ref.~\cite{Adelberger11-RMP}. 

The resulting $S_{34}(0)$ values for each data set are then again fitted together in order to obtain one combined value, again as in Ref.~\cite{Adelberger11-RMP}. The result, based on Refs.~\cite[present]{NaraSingh04-PRL,Bemmerer06-PRL,Brown07-PRC,DiLeva09-PRL,Kontos13-PRC}, is $0.561\pm0.014_{\rm stat}$~keV\,b, with the uncertainty obtained by multiplying the raw uncertainty resulting from the fit with the inflation factor. This can be compared with the "Solar Fusion II" result of $S(0)$ = $(0.56\pm0.02_{\rm stat}\pm0.02_{\rm syst})$~keV\,b \cite{Adelberger11-RMP}. 

In "Solar Fusion II", the systematic uncertainty results from the extrapolation from the energies where many different experiments are available to the solar Gamow peak. For the purposes of BBN, instead of an extrapolation only an interpolation is needed (fig.~\ref{fig:sfactor}). Therefore, this latter error bar can be omitted here.

This result is lower than the previously evaluated value of $S_{34}(0) = 0.580\pm$0.043~keV\,b \cite{Cyburt08-PRC} that has been used in several BBN calculations\cite{Pospelov10-ARNPS,Kusakabe13-PLB,Coc14-JCAP}. When converting to the peak of the BBN sensitivity range, from the present work a value of $S_{34}(226\,{\rm keV}) = 0.485\pm$0.012~keV\,b is found, very close to the previous $0.487\pm$0.036~keV\,b \cite{Cyburt08-PRC} but more precise. 
The increase in precision is due to three factors. First, the adoption of the "Solar Fusion II" approach that gives prominence to the fact that $S_{34}(E)$ has been measured in a number of independent precision experiments, with mutually consistent results. Second, the addition of new data points, including the present one, since 2008. Third, the theory error used in "Solar Fusion II" is not applicable here, as no extrapolation is needed.

\section{BBN reaction rate}
\label{sec:ReactionRate}

\begin{figure}
\centering
\includegraphics[width=\linewidth]{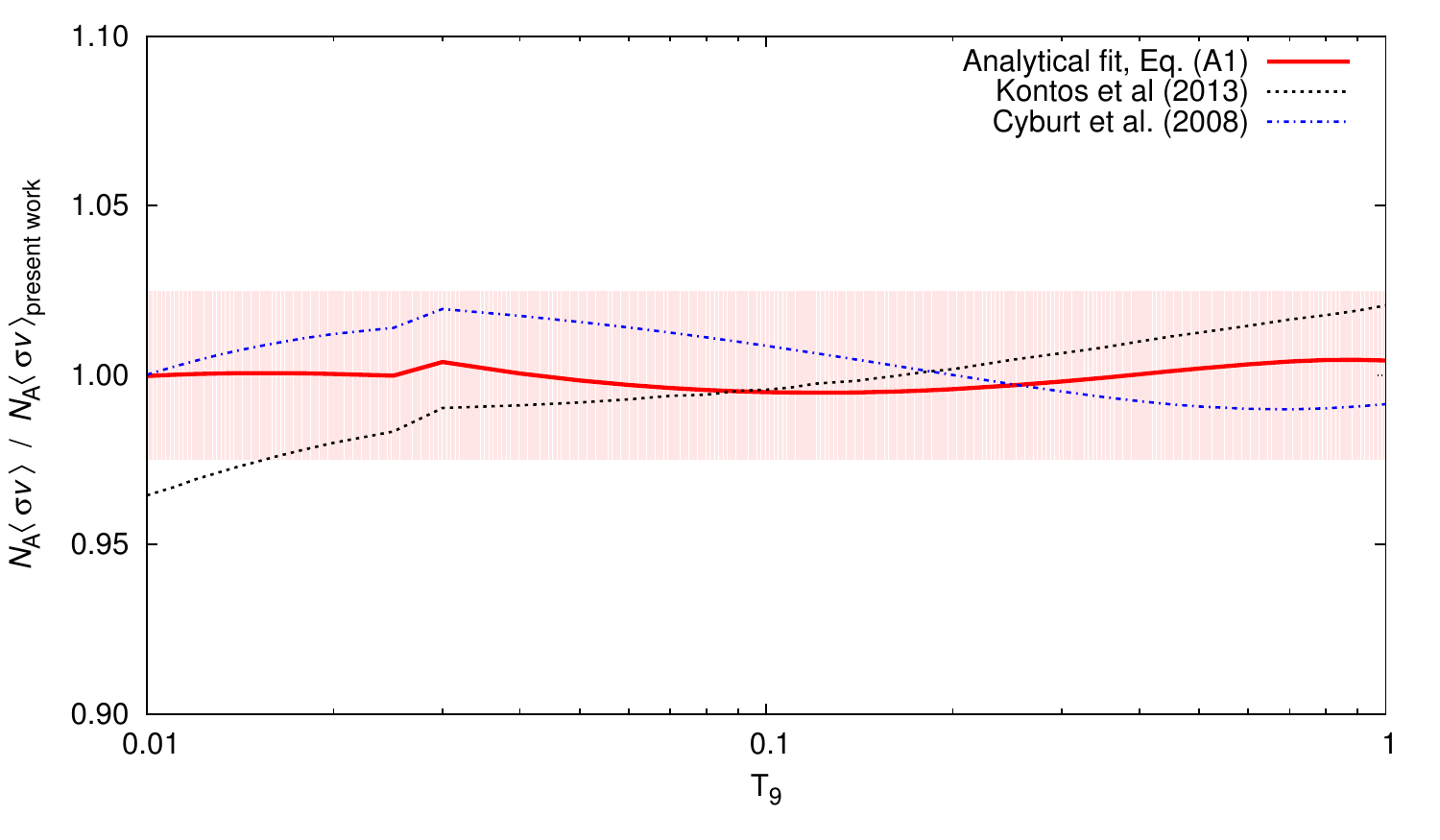}
\caption{Thermonuclear reaction rate for the $^3$He($\alpha,\gamma$)$^7$Be reaction, relative to the present rate, by Kontos {\it et al.} (black dashed curve, \cite{Kontos13-PRC}) and by Cyburt {\it et al.} (blue dot-dashed curve, \cite{Cyburt08-PRC}). The analytical fit function from Eq.~(\ref{eq:tnrrfit}) is also shown (red full curve).}
\label{fig:TNRR}
\end{figure}

The S-factor curve resulting from the combined fit described in the previous section (Fig.~\ref{fig:sfactor}) has subsequently been used to compute the thermonuclear reaction rate $R_{34}$ for its range of applicability, i.e. for $0.001 \leq T_9 \leq 1.0$, which includes the relevant temperature range for BBN (sec.\,\ref{sec:Introduction} and Fig.~\ref{fig:TNRR}). 

For higher temperatures $T_9 > 1.0$, the conclusions depend on the slope of the excitation function above 1\,MeV. Different theoretical papers give different slopes for $E$ $>$ 1\,MeV \cite{Kajino86-NPA,Mertelmeier86-NPA,Nollett01-PRC,Neff11-PRL}. However, this temperature range is irrelevant for BBN (sec.\,\ref{sec:Introduction}) and thus excluded from consideration here. The tabulated reaction rate and an analytical fitting function can be found in the Appendix.

The new rate was then used as input in the PArthENoPE BBN code \cite{Pisanti08-CPC}. Among publicly available codes  \cite{Smith93-ApJS,Pisanti08-CPC}, PArthENoPE incorporates the more recent reaction rate data. For the present purposes, the physics input to PArthENoPE was updated for the following three parameters: $^3$He($\alpha,\gamma$)$^7$Be reaction rate (present work), baryonic density \cite{Planck14-Cosmology} and neutron lifetime \cite{PDG14}. See also sec.~\ref{sec:InputData} for the latter two parameters. 
 
The resulting BBN lithium abundance is
\begin{equation}
^7{\rm Li/H} = 5.0 \times 10^{-10} 
\end{equation}
By repeating the BBN calculation with the upper and lower limits given by the error on $R_{34}$, it is found that the present 2.5\% error for $R_{34}$ contributes just 2.4\% uncertainty to the error budget of $^7{\rm Li/H}$. This value is to be compared with a previous contribution of 5.3\% that can be estimated by using the previous $R_{34}$ error \cite{Cyburt08-PRC} and the previous correlation coefficient \cite{Coc14-JCAP}. The total uncertainty of $^7{\rm Li/H}$ has previously been estimated to be 8\% \cite{Coc14-JCAP}. When subtracting the previous $R_{34}$ contribution in quadrature and adding the present, new $R_{34}$ contribution, a new total relative uncertainty of 6\% can be estimated for $^7{\rm Li/H}$, leading to a final value of $^7{\rm Li/H} = (5.0 \pm 0.3) \times 10^{-10}$. However, this estimated total uncertainty still needs to be borne out in a full BBN calculation re-analyzing in detail also the error budget contributions by parameters other than $R_{34}$.

The present $^7{\rm Li/H}$ value is well above the so-called Spite plateau of lithium abundances \cite{Fields11-ARNPS}, and even further above the lithium values or limits found in extremely metal-poor stars \cite[e.g.]{Caffau11-Nature}.
The recent predicted lithium isotopic ratio \cite{Anders14-PRL} does not change outside the error bar with the present new $^7$Li/H result, it remains $^6$Li/$^7$Li = (1.5$\pm$0.3)$\times10^{-5}$. 
 
\section{Summary and outlook}
\label{sec:Summary}

The astrophysical S-factor of the $^3$He($\alpha,\gamma$)$^7$Be reaction rate has been determined from the measured $^7$Be solar neutrino flux, using the standard solar model. The new data point of $S_{34}^{^7\text{Be}}(23^{+6}_{-5}\,\text{keV})$ = 0.548$\pm$0.054~keV\,b was then used to re-evaluate the excitation function in the energy range relevant for Big Bang nucleosynthesis. A combined average of $S_{34}(0) = 0.561\pm$0.014~keV\,b is found. 

Using the new excitation function, the $^3$He($\alpha,\gamma$)$^7$Be thermonuclear reaction rate was re-computed for the Big Bang energy range, and the fit coefficients for the new recommended rate are given. 

The present results are consistent with, but more precise than previous evaluations. The precision of this solar neutrino based approach will increase even further once the puzzle given by the solar elemental abundances is solved. 

\begin{acknowledgments}
This work was supported by the Helmholtz Association (HGF) through the Nuclear Astrophysics Virtual Institute (NAVI, HGF VH-VI-417), and by the Excellence Initiative of the German Federal and State Governments (TU Dresden Institutional Strategy, program "support the best"). 
\end{acknowledgments}

~


\appendix*

\section{Tabulated values and parameterization of the reaction rate}

The reaction rate (Table~\ref{tab:rates}) is reproduced within $\pm$0.5\% for $0.01 < T_9 < 1.0$ (Fig.~\ref{fig:TNRR}) by the following analytical function:

\begin{eqnarray}\label{eq:tnrrfit}
R_{34} & = & p_{1}{T_{9}}^{-\frac{2}{3}}\exp(p_{2}{T_{9}}^{-\frac{1}{3}})\times  \\  
& & (1+p_{3}T_{9}+p_{4}T_{9}^{2}+p_{5}T_{9}^{3}+p_{6}T_{9}^{4}) \nonumber
\end{eqnarray}

The fit parameters are given in Table~\ref{table:coefficients}.

\begin{table}[b]
\begin{center}
\begin{tabular}{lcrl}
$p_{1}$ &=& 5.497 	&$\times 10^{6}$ \\
$p_{2}$ &=& -1.281 	&$\times 10^{1}$ \\
$p_{3}$ &=& -2.335 	&$\times 10^{-1}$ \\
$p_{4}$ &=& 5.108 	&$\times 10^{-2}$ \\
$p_{5}$ &=& -1.672 	&$\times 10^{-3}$ \\
$p_{6}$ &=& -4.724 	&$\times 10^{-4}$ \\
\end{tabular}
\end{center}
\caption{\label{table:coefficients}
Coefficients for the reaction rate fit in Eq.~\ref{eq:tnrrfit}.}
\end{table}

\begin{table}[b]
\begin{center}
{\small
\begin{tabular}{cccc}
\hline\noalign{\smallskip}
$T_9$ & Reaction rate & $T_9$ & Reaction rate \\
\noalign{\smallskip}
\hline
0.001	&	$	1.339	\times	10^{	-47	}	$	&	0.07	&	$	1.013	\times	10^{	-6	}	$	\\
0.002	&	$	2.475	\times	10^{	-36	}	$	&	0.08	&	$	3.581	\times	10^{	-6	}	$	\\
0.003	&	$	7.147	\times	10^{	-31	}	$	&	0.09	&	$	1.038	\times	10^{	-5	}	$	\\
0.004	&	$	1.975	\times	10^{	-27	}	$	&	0.10	&	$	2.589	\times	10^{	-5	}	$	\\
0.005	&	$	5.518	\times	10^{	-25	}	$	&	0.11	&	$	5.747	\times	10^{	-5	}	$	\\
0.006	&	$	4.040	\times	10^{	-23	}	$	&	0.12	&	$	1.162	\times	10^{	-4	}	$	\\
0.007	&	$	1.243	\times	10^{	-21	}	$	&	0.13	&	$	2.178	\times	10^{	-4	}	$	\\
0.008	&	$	2.096	\times	10^{	-20	}	$	&	0.14	&	$	3.832	\times	10^{	-4	}	$	\\
0.009	&	$	2.279	\times	10^{	-19	}	$	&	0.15	&	$	6.398	\times	10^{	-4	}	$	\\
0.010	&	$	1.778	\times	10^{	-18	}	$	&	0.16	&	$	1.021	\times	10^{	-3	}	$	\\
0.011	&	$	1.071	\times	10^{	-17	}	$	&	0.18	&	$	2.331	\times	10^{	-3	}	$	\\
0.012	&	$	5.240	\times	10^{	-17	}	$	&	0.20	&	$	4.731	\times	10^{	-3	}	$	\\
0.013	&	$	2.166	\times	10^{	-16	}	$	&	0.25	&	$	1.936	\times	10^{	-2	}	$	\\
0.014	&	$	7.789	\times	10^{	-16	}	$	&	0.30	&	$	5.619	\times	10^{	-2	}	$	\\
0.015	&	$	2.490	\times	10^{	-15	}	$	&	0.35	&	$	1.306	\times	10^{	-1	}	$	\\
0.016	&	$	7.203	\times	10^{	-15	}	$	&	0.40	&	$	2.606	\times	10^{	-1	}	$	\\
0.018	&	$	4.712	\times	10^{	-14	}	$	&	0.45	&	$	4.652	\times	10^{	-1	}	$	\\
0.020	&	$	2.372	\times	10^{	-13	}	$	&	0.50	&	$	7.636	\times	10^{	-1	}	$	\\
0.025	&	$	6.018	\times	10^{	-12	}	$	&	0.60	&	$	1.714	\times	10^{	0	}	$	\\
0.03	&	$	7.019	\times	10^{	-11	}	$	&	0.70	&	$	3.243	\times	10^{	0	}	$	\\
0.04	&	$	2.515	\times	10^{	-9	}	$	&	0.80	&	$	5.454	\times	10^{	0	}	$	\\
0.05	&	$	3.177	\times	10^{	-8	}	$	&	0.90	&	$	8.422	\times	10^{	0	}	$	\\
0.06	&	$	2.184	\times	10^{	-7	}	$	&	1.00	&	$	1.220	\times	10^{	1	}	$	\\
\hline
\end{tabular}
}
\end{center}
\caption{\label{tab:rates}
$^3$He($\alpha,\gamma$)$^7$Be reaction rate in cm$^{3}$s$^{-'1}$mole$^{-ˆ'1}$. }
\end{table} 


\end{document}